\documentstyle[preprint,epsfig,eqsecnum,aps]{revtex}
\title{Microcanonical studies concerning the recent experimental
  evaluations of the nuclear caloric curve}
\author{Al. H. Raduta and Ad. R. Raduta}
\address{National Institute of Physics and Nuclear Engineering,\\ 
  Bucharest, POB MG-6, Romania}
\begin{document}
\maketitle
\begin{abstract}
The microcanonical multifragmentation model from [Al. H. Raduta and 
Ad. R. Raduta, Phys. Rev. C {\bf 55}, 1344 (1997); {\bf 56}, 2059
(1997); {\bf 59}, 323 (1999)] is refined and improved by taking into 
account the experimental
discrete levels for fragments with $A \le 6$ and by including the 
stage of sequential decay of the primary excited fragments. The caloric curve 
is reevaluated and the heat capacity at constant volume curve 
is represented as a function of excitation energy and temperature. The
sequence of equilibrated sources formed in the reactions studied by the ALADIN
group ($^{197}$Au+$^{197}$Au at 600, 800 and 1000 MeV/nucleon bombarding 
energy) is deduced by fitting simultaneously the model predicted mean
multiplicity of intermediate mass fragments ($M_{IMF}$) and charge  
asymmetry of the two largest fragments ($a_{12}$) versus 
bound charge ($Z_{bound}$) on the corresponding experimental
data. Calculated HeLi isotopic temperature curves as a function of the bound
charge are compared with the experimentally deduced ones.
\end{abstract}

\newpage

\section{Introduction}
Nuclear multifragmentation is presently intensely studied both
theoretically and experimentally. Due to the similitude existent between the
nucleon-nucleon interaction and the van der Waals forces, signs of 
a liquid-gas phase transition in nuclear matter are searched. 
While the theoretical calculations concerning this problem started at the
beginning of 1980 \cite{Ran}, the first experimental evaluation of the nuclear
caloric curve was reported in 1995 by the ALADIN group \cite{Pocho}. A wide
plateau situated at around 5 MeV temperature lasting from 3 to 10 MeV/nucleon
excitation energy was identified. The fact was obviously associated with the
possible existence of a liquid-gas phase transition in nuclear matter and
generated new motivations for further theoretical and experimental
work. Similar experiments of EOS \cite{Eos} and INDRA \cite{Indra} followed
shortly. Using different reactions they obtained slightly different caloric
curves, the plateau - like region being absent in the majority of cases. 
Factors contributing to these discrepancies are both the precision
of the experimental measurements and the finite-size effects of the caloric
curve manifested through the dependency of the 
equilibrated sources [$E^*(A)$] sequence on the reaction type.

Concerning the first point of view, recent reevaluations of the ALADIN group
concerning the kinetic energies of the  emitted neutrons brought corrections of
about 10 $\%$ (in the case of the reaction $^{197}$Au+$^{197}$Au, 
600 MeV/nucleon). More importantly however it was proven that the energies of
the spectator parts are growing with approximately 30 $\%$ in the bombarding 
energy
interval 600 to 1000 MeV/nucleon. On the other side, the universality of the
quantity $M_{IMF}(Z_{bound})$ subject to the bombarding energy 
variation (which was
theoretically proven \cite{Schuttauf,Pocho_ppnp} to be a signature of
statistical equilibrium) suggests that for the above-mentioned reactions the
equilibrated sources sequence [$E^*(A)$] should be the same. 
Consequently, we deal with an important nonequilibrium part included in the
measured source excitation energies which may belong to both pre-equilibrium or
pre-break-up stages \cite{Trautmann}. The SMM 
calculations suggest a significant quantity of nonequilibrium energy even in 
the case of the 600 MeV/nucleon bombarding energy reaction 
\cite{Trautmann,Xi,Mueller-Hir}.

Thus, the necessity of accurate theoretical descriptions of the break-up 
stage and of the sequential secondary particle emission appears to be
imperative in order to distinguish between the equilibrium and nonequilibrium
parts of the measured excitation energies. These approaches should strictly
obey the constrains of the physical system which, in the case of nuclear
multifragmentation, are purely microcanonic. As we previously underlined
\cite{Temp1,Temp2}, in spite of their success in reproducing some 
experimental data, the two widely used statistical multifragmentation models 
(SMM \cite{Bondorf} and MMMC \cite{Gross}) are not strictly satisfying the
microcanonical rules.

The present paper describes some refinements and improvements 
brought to the sharp microcanonical multifragmentation model proposed in 
\cite{Noi1,Noi2} and also the employment of the model in its new version in the
interpretation of the recent experimental data of the ALADIN group
\cite{Trautmann,Xi}.

The improvements brought to the model \cite{Noi1,Noi2} are 
presented in Section II. Section III  presents the new evaluations of
temperature curves and the first evaluations (performed with this model) 
of heat
capacities at constant volume ($C_V$) represented as a function of system
excitation energy and temperature and also the comparison between the model
predictions and the recent experimental HeLi isotopic temperature curve 
[$T_{HeLi}(Z_{bound})$] \cite{Trautmann,Xi}. Conclusions are drawn in Section
IV.  

\section{Improvements brought to the microcanonical multifragmentation model}

The improvements brought to the microcanonical multifragmentation model
concerns both the {\it break-up} stage and the 
{\it secondary particle emission} stage.\\
(i) {\it Primary break-up refinements}\\
Comparing to the version of Ref.\cite{Noi1,Noi2,Temp1} the present model has
the following new features:\\
(a) The experimental discrete energy 
levels are replacing the level density for fragments
with $A\le 6$ (in the previous version of the model a Thomas Fermi type level
density formula was used for all particle excited states). In this respect, 
in the statistical weight of a configuration
and the correction factor formulas \cite{Noi1,Noi2} the level density
functions are replaced by the degeneracies of the discrete levels, $(2S_i+1)$
(here $S_i$ denotes the spin of the $i$th excited level). As a criterion 
for level selection (i.e. the level life-time must be greater than the typical
time of a fragmentation event) we used $\Gamma \le$ 1 MeV, where $\Gamma$
is the width of the energy level.\\
(b) In the case of the fragments with $A >6 $ the level density formula is
modified as to take into account the strong decrease of the fragments 
excited states life-time (reported to the standard duration of a 
fragmentation event) with the increase of their excitation energy. To this
aim the Thomas Fermi type formula \cite{Noi1} is completed with the 
factor $\exp(-\epsilon/\tau)$ (see Ref.\cite{Randrup}):
\begin{equation}
  \rho(\epsilon)=\frac 1{\epsilon\sqrt{48}}\exp{(2\sqrt{a\epsilon})}
  \exp{(-\epsilon/\tau)},
\end{equation}
where $a=A/\alpha$, $\alpha=4.7(1.625+\epsilon/B(A,Z))$ and $\tau=9$.\\
(ii) {\it Inclusion of the secondary decay stage}\\
For the $A>6$ nuclei it was observed that the fragments excitation energies
are sufficiently small such as the sequential evaporation scheme is perfectly
applicable. According to Weisskopf theory \cite{Weisskopf} (extended as to
account for particles larger than $\alpha$), the probability of emitting a
particle $j$ from an excited nucleus is proportional to the quantity:
\begin{equation}
  W_j=\sum\limits_{i=0}^n \int_0^{E^*-B_j-\epsilon_i^j}
  \frac{g_i^j\mu_j\sigma_j(E)}{\pi^2\hbar^3}~
  \frac{\rho_j(E^*-B_j-\epsilon_i^j-E)}{\rho(E^*)}E{\text d}E,
\end{equation}
where $\epsilon_i$ are the stable excited states of the fragment $j$ subject
to particle emission (their upper limit is generally around 7 - 8 MeV), 
$E$ is the kinetic energy of the formed pair in the center of mass (c.m.)
frame, $g_i^j=2S_i+1$ is the degeneracy of the level $i$,
$\mu_j$ and $B_j$ are respectively the reduced mass of the pair and the
separation energy of the particle $j$ and finally $\sigma_j$ 
is the inverse reaction cross-section. Due to the specificity of the
multifragmentation calculations we considered the range of the emitted 
fragments $j$ up to the $A=16$ limit.
For the inverse reaction cross-section we have used the optical model based
parametrization from Ref. \cite{Botvina}. The sequential evaporation process
is simulated by means of standard Monte Carlo (see for example 
\cite{Dostrovsky}).

For nuclei with $4\le A\le 6$ (the only excited states of $A=4$ nuclei taken
into consideration are few states higher than 20 MeV belonging to the
$\alpha$ particle) depending on their amount of excitation we consider 
{\it secondary break-up} for $\epsilon > B(A,Z)/3$ and Weisskopf evaporation 
otherwise (here $\epsilon$ is the excitation energy of the fragment $(A,Z)$
and $B(A,Z)$ is its binding energy). 
The microcanonical weight formulas have the usual form \cite{Noi1}  
excepting the level density functions which are here replaced by the discrete
levels degeneracies. Due to the reduced dimensions of the $A<6$ systems, the
break-up channels are countable (and a classical Monte Carlo simulation is
appropriate) when a mean field approach is used
for the Coulomb interaction energy.
In this respect, the Wigner-Seitz approach \cite{Bondorf} is employed for the
Coulomb interaction:
\begin{equation}
  U_C=\frac35 \ \frac{Z_0^2 e^2}{R}-\sum\limits_i \frac35 \ 
  \frac{Z_i^2 e^2}{R_{A_iZ_i}^C},
\end{equation}
where $A$ and $Z$ denotes the mass and the charge of the source nucleus, the
resulting fragments have the index $i$, $R_{A_iZ_i}^C/R_{A_iZ_i}=(1+\kappa)
^{1/3}\left[(Z_i/A_i)/(Z/A)\right]^{1/3}$
and $V=(1+\kappa) V_0$. Here $V$ denotes the break-up volume and $V_0$ the
volume of the nucleus at normal density. It should be added that
 $R$ is the radius of the source
nucleus at break-up and $R_{A_iZ_i}$ is the radius of fragment $i$ at
normal density.

For each event of the primary break-up simulation, the entire chain of 
evaporation and secondary break-up events is Monte Carlo simulated.

\section{Results}

Using the improved version of the microcanonical multifragmentation model, the
caloric curves corresponding to two freeze-out radii (R=2.25 A$^{1/3}$ and 
R=2.50 A$^{1/3}$ fm) are reevaluated for the case of the source nucleus
(70, 32) (the microcanonical caloric curves evaluated with the initial version
of the model are given in Ref. \cite{Temp1,Temp2,Temp3}). These are presented 
in
Fig. 1 (a). One can observe that the main features of the caloric curve from
Refs. \cite{Temp1,Temp2,Temp3} are reobtained. Thus, one can recognize the 
liquid-like 
region at the beginning of the caloric curve, then a large plateau-like 
region and finally the linearly increasing gas-like region.
One may also notice that the caloric curve behavior at the freeze-out radius
variation is maintained: The decrease of the freeze-out radius leads to
a global lifting of the caloric curve.

As it is well known, the curves of the constant volume heat capacity ($C_V$) 
as a function of system excitation energy ($E^*$) 
and as a function of temperature ($T$) may
provide important information concerning the transition region and the
transition order. For this reason the curves $C_V(E^*)$ and $C_V(T)$ have been
evaluated (see Fig. 1 (a) and Fig. 1 (b)). We remind that the
constant volume heat capacity ($C_V$) is calculable in the present model using
the formula \cite{Temp2}:
\begin{equation}
  C_V^{-1}=1-T^2\left<\left[ \left(\frac32 N_C- \frac52\right)\frac1K
    \right]^2\right>+T^2\left<\left(\frac32 N_C- \frac52\right)
    \frac1{K^2}\right>.
\end{equation}
It can be observed that the $C_V(E^*)$ curve has a sharp maximum around 4.5
MeV/nucleon excitation energy for both considered freeze-out radii. This
suggests that a phase transition exists in that region.
The transition temperatures can be very well distinguished by analyzing the 
$C_V(T)$. One can observe two sharp-peaked maxima pointing the transition
temperatures corresponding to the two considered freeze-out radii.

In order to make a direct comparison between the calculated HeLi isotopic
temperature and the recent experimental results 
\cite{Trautmann,Xi,Mueller-Hir} one has to deduce the
sequence of excitation energy as a function of the system dimension
[$E^*(A)$]. This is done as in Refs. \cite{Trautmann,Xi} using as matching 
criterion the
simultaneously reproduction of the  
$\left<M_{IMF}\right>(\left<Z_{bound}\right>)$ and
$\left<a_{12}\right>(\left<Z_{bound}\right>)$ curves. 
This couple of curves can fairly well identify the
dimension and the excitation of the equilibrated nuclear source 
\cite{Trautmann,Xi}.
Here $M_{IMF}$ stands for the multiplicity of
intermediate mass fragments and is defined as the number of fragments with 
$3\le Z\le 30$ from a fragmentation event while 
$a_{12}$ denotes the charge asymmetry of the two largest fragments 
and, for one fragmentation event is defined
as $a_{12}=(Z_{max}-Z_2)/(Z_{max}+Z_2)$ with $Z_{max} \le Z_2 \le 2$ where
$Z_{max}$ is the maximum charge of a fragment and $Z_2$ is the second largest 
charge of a fragment in the respective event. $Z_{bound}$ represents the 
{\it bound charge} in one fragmentation event and is defined as the sum 
of the charges of all fragments with $Z\ge 2$.

The simultaneous fit of the calculated curves 
$\left<M_{IMF}\right>(\left<Z_{bound}\right>)$ and 
$\left<a_{12}\right>(\left<Z_{bound}\right>)$ on the
corresponding experimental data ($^{197}$Au+$^{197}$Au at 1000 MeV/nucleon)
is given in Fig. 2. The agreement is very
good. The equilibrated source sequence [$E^*(A)$] we used for this purpose 
is given in Fig. 3 together with the experimental evaluations of the
excitation energies as a function of source dimension for the reaction
$^{197}$Au+$^{197}$Au at 600, 800 and 1000 MeV/nucleon. The theoretically
obtained sequence is relatively close to the experimental line corresponding to
 600 MeV/nucleon bombarding energy.
The deviations between the calculated equilibrated source sequence and the 
three experimental lines suggest that the experimental evaluations
contain a quantity of non-equilibrium energy which grows with increasing the
bombarding energy. As suggested in Ref. \cite{Trautmann,Xi}, 
its origin may be situated in both the pre-equilibrium and pre-break-up stage.
 These deviations are exclusively due to the neutron
kinetic energies which, reevaluated \cite{Trautmann,Xi}
 from the 1995 data \cite{Pocho}, are much larger. 

It should also be pointed that apart from the SMM predictions
\cite{Trautmann,Xi,Mueller-Hir}, the quantity of non-equilibrium energy
predicted by the present model is smaller and thus 
the model predicted equilibrated source sequence is closer to the
experimental line of the 600 MeV/nucleon bombarding energy reaction.

After evaluating the  sequence of the equilibrated sources a direct comparison
the HeLi calculated isotopic temperature curve with the ones recently
evaluated by the ALADIN group \cite{Trautmann,Xi} is performed. 
To this purpose the 
uncorrected Albergo temperature is used: 
 $T_{HeLi}=13.33/\ln\left[2.18~\left(Y_{^6Li}/Y_{^7Li}\right)
/\left(Y_{^3He}/Y_{^4He}\right)\right]$, the experimental 
predictions being divided by $f_T=1.2$ (which is  the factor used in
the ALADIN evaluation of the HeLi caloric curve chosen as to average the 
QSM, GEMINI and MMMC models predictions). The result is represented in Fig. 4
as a function of $Z_{bound}$. 
It can be observed that the agreement between the calculated 
$T_{HeLi}(Z_{bound})$ and the experimental data 
corresponding to the
$^{197}$Au+$^{197}$Au reaction at 600 and 1000 MeV/nucleon bombarding 
energy is excellent on the entire range of $Z_{bound}$.
In comparison, the SMM model predicts in the region 
$Z_{bound}\le 25$ a
curve steeper than the experimental data.

\section{Conclusions}

Sumarizing, the microcanonical multifragmentation model from Ref. 
\cite{Noi1,Temp1} is improved by refining the primary break-up part
and by including the secondary particle emission. The caloric curve
rededuced with the new version of the model preserves its general aspect
\cite{Temp1,Temp2,Temp3} manifesting an important plateau-like
region. The transition regions are clearly indicated by the sharp maxima of the
$C_V(E^*)$ and $C_V(T)$ curves. The model proves the ability of simultaneously 
fitting the "definitory" characteristics of the nuclear multifragmentation
phenomenon $\left<M_{IMF}\right>(\left<Z_{bound}\right>)$ and 
$\left<a_{12}\right>(\left<Z_{bound}\right>)$. Evaluating 
the equilibrated source sequence $E^*(A)$ [by using the criterion of
reproducing both $\left<M_{IMF}\right>$ and $\left<a_{12}\right>$ versus 
$\left<Z_{bound}\right>$], a
nonequilibrium part of the experimentally evaluated excitation energy growing
with the increase of the bombarding energy is identified.
 The direct comparison of the calculated HeLi caloric curve shows an excellent
agreement with the experimental HeLi curves recently evaluated by the ALADIN
group. 

\begin{figure}
  \begin{center}
  \epsfig{file=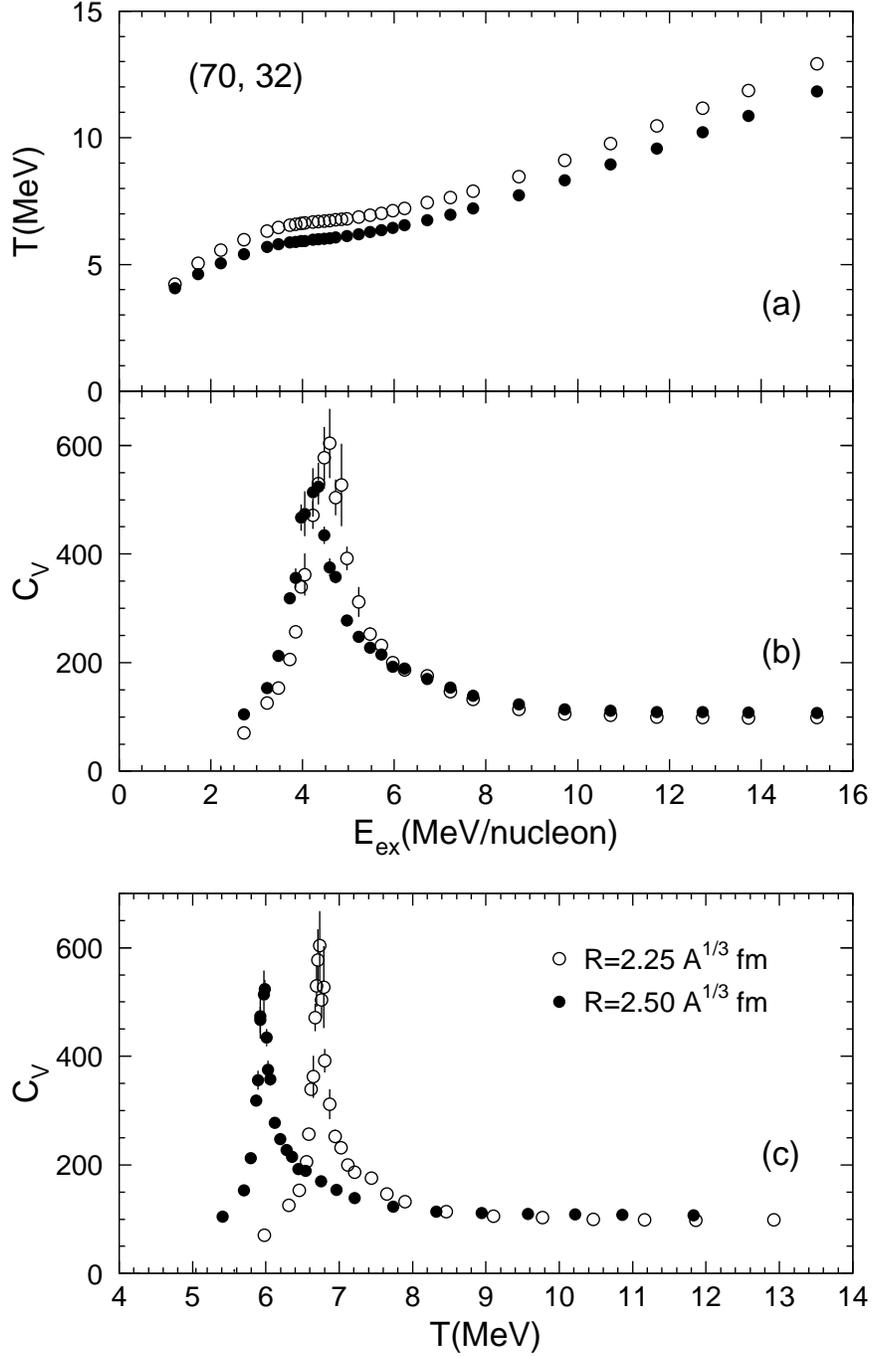,height=18cm,angle=0}
  \end{center}
  \caption{ 
    Microcanonical temperature as a function of source excitation energy (a), 
    heat capacity at constant volume as a function of source excitation energy
    (b) heat capacity at constant volume as a function of microcanonical
    temperature (c). Calculations have been performed for the source nucleus
    (70, 32) with two values of the freeze-out radius.}
\end{figure}

\begin{figure}
  \begin{center}
  \epsfig{file=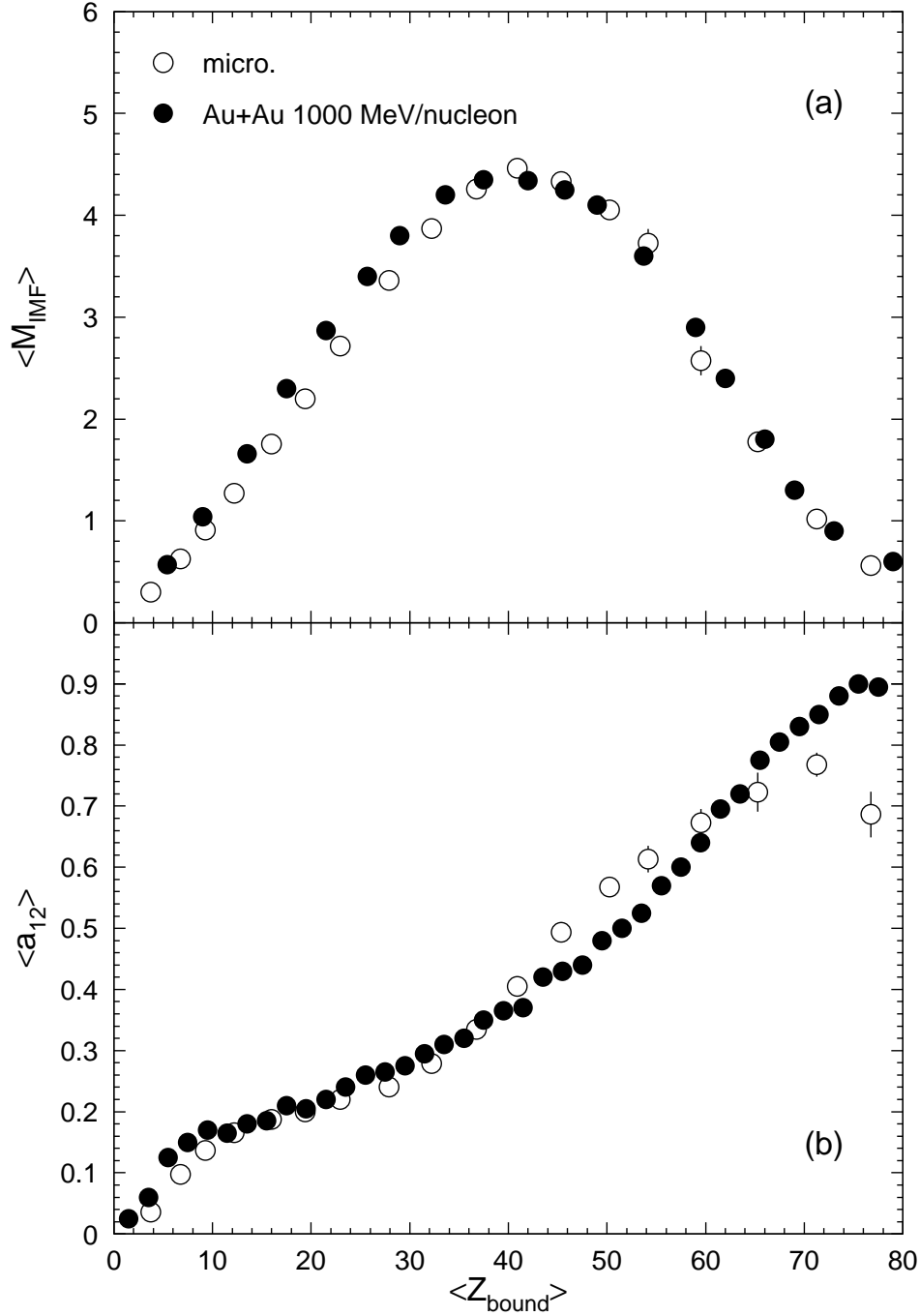,height=18cm,angle=0}
  \end{center}
  \caption{$\left<M_{IMF}\right>(\left<Z_{bound}\right>)$ and 
    $\left<a_{12}\right>(\left<Z_{bound}\right>)$ evaluated by means of the 
    microcanonical model in comparison with the experimental evaluations
    corresponding to the reaction $^{197}$Au+$^{197}$Au at 1000 MeV/nucleon
    bombarding energy. (The deviation of calculated $\left<a_{12}\right>$ 
    from experimental data for $Z_{bound}\ge 65$ is, as explained in 
    [7], due to some detection problems.)}
\end{figure}

\begin{figure}
  \begin{center}
  \epsfig{file=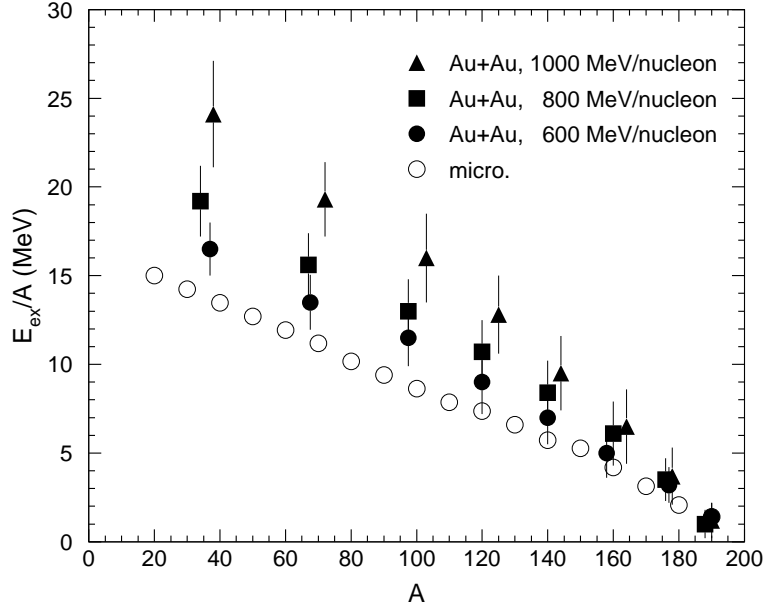,height=8cm,angle=0}
  \end{center}
  \caption{The sequence of equilibrated sources evaluated by means of 
    the microcanonical model (open circles) 
    following the criterion of simultaneous fitting of
    the calculated $M_{IMF}(Z_{bound})$ and $a_{12}(Z_{bound})$ on the
    corresponding experimental data. The close symbols represent the sequence
    of excitation energy experimentally measured for the reaction 
    $^{197}$Au+$^{197}$Au at 600, 700, 1000 MeV/nucleon bombarding energy [7].}
\end{figure}

\begin{figure}
  \begin{center}
  \epsfig{file=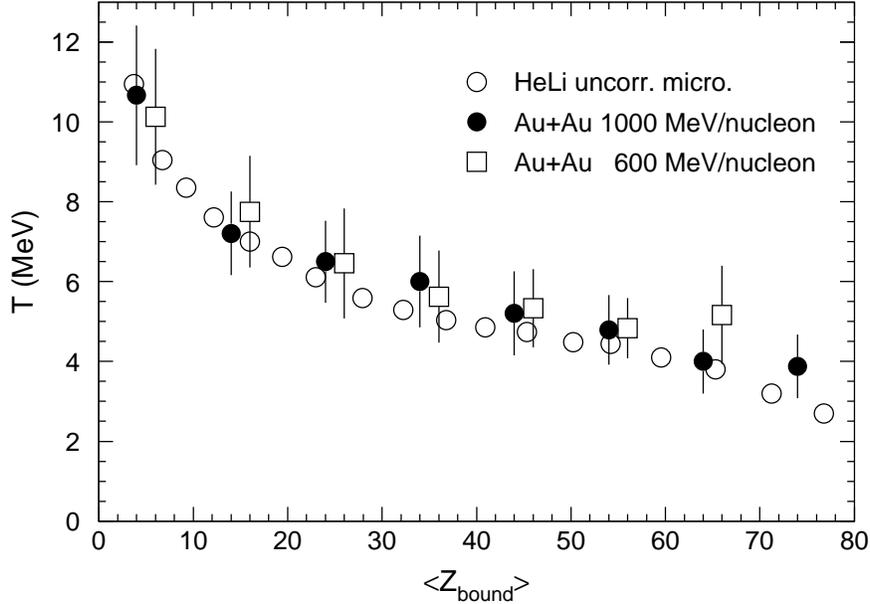,height=8cm,angle=0}
  \end{center}
  \caption{
    HeLi temperature curves evaluated  with the microcanonical model (open
    circles) in comparison with the experimental HeLi temperatures [7,8]
    corresponding to the reactions $^{197}$Au+$^{197}$Au at 600 and 1000 
    MeV/nucleon bombarding energy.}
\end{figure}

\end{document}